\documentclass[groupedaddress,
  aps,
  pra,
  amsmath,amssymb,
  reprint,
]{revtex4-1}
\usepackage{graphicx,color}
\usepackage{amsmath}
\usepackage{natbib}
\usepackage{epsfig}
\begin{document}

\title{\color{blue} Elastic properties of dense hard sphere fluids}

\author{Sergey Khrapak}
\email{Sergey.Khrapak@dlr.de}
\affiliation{Institut f\"ur Materialphysik im Weltraum, Deutsches Zentrum f\"ur Luft- und Raumfahrt (DLR), 82234 We{\ss}ling, Germany;
Joint Institute for High Temperatures, Russian Academy of Sciences, 125412 Moscow, Russia}

\begin{abstract}
A new analysis of elastic properties of dense hard sphere (HS) fluids is presented, based on the expressions derived by Miller [J. Chem. Phys. {\bf 50}, 2733 (1969)].  Important consequences for HS fluids in terms of sound waves propagation, Poisson's ratio, Stokes-Einstein relation, and generalized Cauchy identity are explored.  Conventional expressions for high-frequency elastic moduli for simple systems with continuous and differentiable interatomic interaction potentials are known to diverge when approaching the HS repulsive limit. The origin of this divergence is identified here. It is demonstrated that these conventional expressions are only applicable for sufficiently soft interactions and should not be applied to HS systems. The reported results can be of interest in the context of statistical physics, physics of fluids, soft condensed matter, and granular materials.  
\end{abstract}

\date{\today}

\maketitle

\section{Introduction}

The behavior of elastic moduli in systems with steeply repulsive interaction potentials, when approaching the hard-sphere (HS) limit, appears as a controversial issue. The conventional expressions for the high-frequency (instantaneous) bulk and shear moduli~\cite{ZwanzigJCP1965,Schofield1966} predict their divergence as the HS limit is approached from the side of soft interactions~\cite{Frisch1966,HeyesJCP1994}. Various aspects of this divergence have been discussed in the literature. However, the divergence itself is not consistent with several other observations. For example, elastic moduli of HS solids are well defined~\cite{FrenkelPRL1987,RungePRA1987,LairdJCP1992}. The same applies to the HS glass~\cite{LowenJPCM1990}. The isothermal and adiabatic sound velocities of HS fluids follows directly from the equation of state and, thus, the corresponding bulk moduli are also well defined~\cite{RosenfeldJPCM1999}. Dense HS fluids can support both longitudinal and transverse collective excitations~\cite{BrykJCP2017}. indicating that the instantaneous shear modulus is also finite. In a recent study of dense fluids with inverse-power-law (IPL) interactions~\cite{KhrapakSciRep2017} it was observed that the measured longitudinal sound velocity smoothly approaches the HS limiting value as the potential steepness (IPL exponent) increases. All this points towards finite values of the elastic constants on approaching and in the HS limit. This seems physically reasonable: since structural and thermodynamic properties are known to approach smoothly and continuously the HS limit it is unclear why different behaviour should be expected from elastic properties.    

About 50 years ago Miller derived expressions for the shear and bulk moduli of a HS fluid and demonstrated that these are non-singular and well defined~\cite{MillerJCP1969}. For some reasons, this result has not received due attention and was not analyzed at all. After finding that elastic moduli of true HS fluids are finite and well defined, Miller stated, in view of the apparent divergence of conventional expressions in the HS limit, that ``This result may lead one to suspect that a fluid with a highly repulsive, but continuous and differentiable, intermolecular potential may not always be even qualitatively represented by a fluid consisting of perfectly rigid spheres''~\cite{MillerJCP1969}. 

To avoid any misunderstanding, two type of divergences can occur in HS systems. The first is natural and physically transparent: The HS pressure and elastic moduli will diverge upon compression when either the maximally dense fcc close-packing is reached (slow compression) or jamming transition occurs (rapid compression)~\cite{RintoulPRL1996,TorquatoRMP2010}. This type of divergence is not addressed. The divergence which is considered here occurs at a {\it fixed} fluid packing fraction when the interaction potential approaches the limit of infinite repulsion.

The aim of the present work is twofold. First, the missing numerical analysis of Miller's expressions for the HS moduli in three dimensions is provided. In particular, the elastic moduli are expressed in terms of the HS packing fraction. Then, the longitudinal and transverse sound velocities are evaluated in the dense fluid regime and some consequences for Poisson's ratio, Stokes-Einstein relation, and Cauchy identity are discussed. Second, the paradoxical divergence of the conventional elastic moduli expressions on approaching the HS limit is explained. By way of a simple derivation of the expressions for the instantaneous bulk moduli for one-dimensional and three-dimensionsl HS fluids, we demonstrate where exactly a problem arises in the HS limit. It is concluded that the divergence is artificial and the conventional expressions are simply not applicable in the HS limit.           

Elastic moduli play extremely important role in various aspects of condensed matter physics and materials science~\cite{DyreRMP2006,NemilovJNCS2006}. One relevant example from the studies of glass-forming liquids is the shoving model, which relates the alpha-relaxation time to the instantaneous bulk and shear moduli and demonstrates that the shear modulus provides dominant contribution (phenomenon known as ``shear dominance'')~\cite{DyrePRE2004,DyreJCP2012}. Another example comes from the celebrated Lindemann criterion of melting~\cite{Lindemann}. This criterion states that a solid melts when the root-mean-square vibration amplitude of atoms around their equilibrium position reaches a certain fraction ($\sim 0.1$) of the interatomic distance. The vibration amplitude itself is related to the shear and bulk moduli (and the shear modulus again provides dominant contribution)~\cite{BuchenauPRE2014} .  One more example is related to the theory of melting in two spatial dimensions (2D).  According to the Berezinskii-Kosterlitz-Thouless-Halperin-Nelson-Young (BKTHNY) theory~\cite{KosterlitzRMP2017}, 2D melting is a two-stage process. The crystal first melts by dislocation unbinding to an anisotropic hexatic fluid and then undergoes a continuous transition into isotropic fluid. The condition for dislocation unbinding can be expressed in terms of (2D) shear and bulk moduli and shear contribution again dominates~\cite{KhrapakJCP2018}. Since the real physical systems of interest can be characterized by quite different interactions, including sufficiently steep potentials, it is not unimportant to understand what happens with elastic moduli when the HS limit is approached.

\section{Miller's result}\label{MillerRes}

Miller calculated the elastic bulk and shear moduli of the single-component fluid of rigid spheres~\cite{MillerJCP1969}. His method consisted of finding the general form of the stress tensor for a HS system, asserting the assumptions of local equilibrium, end expanding the stress tensor in terms of strains using the technique developed by Green. In the derivation he assumed that the duration of the collision between the hard spheres is zero and that simultaneous triple and higher multiplicity collisions are absent. Only the terms independent of the wavelength were retained in the long-wavelength expansion of the stress tensor. The derivation is quite involved and further details  are not reported here.

Expressed in conventional notation, Miller's result for the shear modulus reads
\begin{equation}\label{G}
G = nT\left[1-\frac{8}{5}\phi g'(1) \right],
\end{equation} 
where $n$ is the density of $N$ particles occupying volume $V$ ($n=N/V$),  $T$ is the temperature (in energy units, so that $k_{\rm B}=1$), $\phi = (\pi/6)n\sigma^3$ is the packing fraction of rigid spheres having a diameter $\sigma$ (or reduced density), $g(x)$ is the radial distribution function ($x=r/\sigma$), and $g'(x)=dg(x)/dx$. As usually, the derivative at contact should be taken from above, that is $g'(1)=\lim_{\varepsilon\rightarrow 0}[d g(x)/dx]_{x=1+\varepsilon}$. The bulk modulus $K$ is then related to the shear modulus $G$ via
\begin{equation}\label{K}
K=2P-\frac{8}{3}nT+\frac{2}{3}\frac{P^2}{nT}+\frac{5}{3}G,
\end{equation}  
where $P$ is the pressure. Originally, only the excess (configurational) contribution to the shear modulus was retained in Ref.~\cite{MillerJCP1969} We simply added the kinetic term $nT$ to this original result to produce Eq.~(\ref{G}) and modified the expression for the bulk modulus (\ref{K}) accordingly. Note that in the low-density limit, the ideal-gas results $K=(5/3)nT$ and $G=nT$ are recovered.      

No further analysis of Eqs.~(\ref{G}) and (\ref{K}) was originally performed, except that the conventional expressions~\cite{ZwanzigJCP1965,Schofield1966} for high-frequency moduli $G$ and $K$ were provided for forthcoming comparisons. The conventional moduli are expressed in terms of the pairwise interaction potential $\varphi (r)$ and the radial distribution function (RDF) $g(r)$. The expressions are provided below for completeness:  
\begin{equation}\label{G_}
G=nT+\frac{2\pi n^2}{15}\int_0^{\infty}dr r^3 g(r)\left[r \varphi''(r)+4\varphi'(r)\right],
\end{equation}
\begin{equation}\label{K_}
K=\frac{5}{3}nT+\frac{2\pi n^2}{9}\int_0^{\infty}dr r^3g(r)\left[r \varphi''(r)-2\varphi'(r)\right].
\end{equation}
The first terms correspond to the kinetic (ideal gas) contribution, the second terms are the excess (configurational) parts.

In order to make any progress with Eqs.~(\ref{G}) and (\ref{K}) we need to specify the derivative of the radial distribution function (RDF)  at contact, $g'(1)$, as well as the proper equation of state $P(n,T)$. For the RDF, the simplest approach is to use the results of Thiele~\cite{ThieleJCP1963} obtained within the framework of the Percus-Yevick (PY) approximate integral equation of state~\cite{ThieleJCP1963,WertheimPRL1963}. This yields (see Appendix for details)     
\begin{equation}\label{RDF}
g'(1) = -\frac{9\phi(1+\phi)}{2(1-\phi)^3}.
\end{equation}
However, the PY result is known to underestimate the absolute magnitude of the derivative. An alternative semi-empirical expression, as simple as the PY approximation (\ref{RDF}), but predicting higher values for $|g'(1)|$ at high densities was therefore proposed~\cite{TaoPRA1992}
\begin{equation}\label{RDF1}
g'(1) = -\frac{9\phi(1+\phi)}{2(1-\phi)^4}.
\end{equation}
Later, molecular dynamics (MD) simulation results for $g'(1)$ have been shown to lie somewhere in between the predictions of formulas (\ref{RDF}) and (\ref{RDF1})~\cite{SigurgeirssonMolPhys2003}. In view of this and having no better recipe, we suggest to take a simple average of Eqs.~(\ref{RDF}) and (\ref{RDF1}), which yields
\begin{equation}\label{RDF2}
g'(1) = -\frac{9\phi(1+\phi)(2-\phi)}{4(1-\phi)^4}.
\end{equation} 
The shear modulus then becomes 
\begin{equation}
G=nT\left[1+\frac{18\phi^2(1+\phi)(2-\phi)}{5(1-\phi)^4}\right].
\end{equation}

Regarding the equation of state, we adopt the well known Carnahan-Starling (CS) formula~\cite{CarnahanJCP1969}. This formula was obtained by postulating that reduced virial coefficients in the virial expansion for HS pressure can be approximated by integers and finding a recursive relation for this coefficients. Although this is an approximation, it provides accurate enough results across the fluid density range and is convenient for practical applications.  If the pressure is written in the form $P(n,T)=nTZ(\phi)$, then the CS compressibility factor is
\begin{equation}
Z(\phi)=\frac{1+\phi+\phi^2-\phi^3}{(1-\phi)^3}.
\end{equation}         

Quantitative results for the instantaneous bulk modulus (\ref{K}) can be compared with those for the adiabatic bulk modulus. The adiabatic bulk modulus is defined as $K_{S}=-V(\partial P/\partial V)_S\equiv n \gamma (\partial P/\partial n)_T$, where $V$ is the volume containing $N$ particles (so that $n=N/V$) and $\gamma=C_p/C_v$ is the adiabatic index. The differentiation is easily performed by noting that $n(\partial/\partial n)=\phi(\partial /\partial \phi)$. The result is~\cite{RosenfeldJPCM1999,KhrapakJCP2016} 
\begin{equation}\label{Adiabatic}
K_{S}=nT\left[Z(\phi)+\phi dZ(\phi)/d\phi+\tfrac{2}{3}Z^2(\phi)\right].
\end{equation}
The first two terms on the right side correspond to isothermal modulus ($\gamma=1$), the last term reflects the adiabatic character of the considered process. In general, the inequality $K\geq K_S$ holds~\cite{Schofield1966}. Based on our previous experience with strongly coupled fluids with isotropic purely repulsive interactions,  an approximate equality $K\simeq K_S$ can be expected~\cite{KhrapakPRE2015,KhrapakPoP2016,KhrapakPRE2018}. In particular, in a recent study of fluids with IPL potentials ($\propto r^{-\ell}$), this approximate relation has been verified in the entire region where the conventional expressions for the elastic moduli are applicable ($\ell\lesssim 25$)~\cite{KhrapakSciRep2017}.    

\begin{figure}
\includegraphics[width=7.5cm]{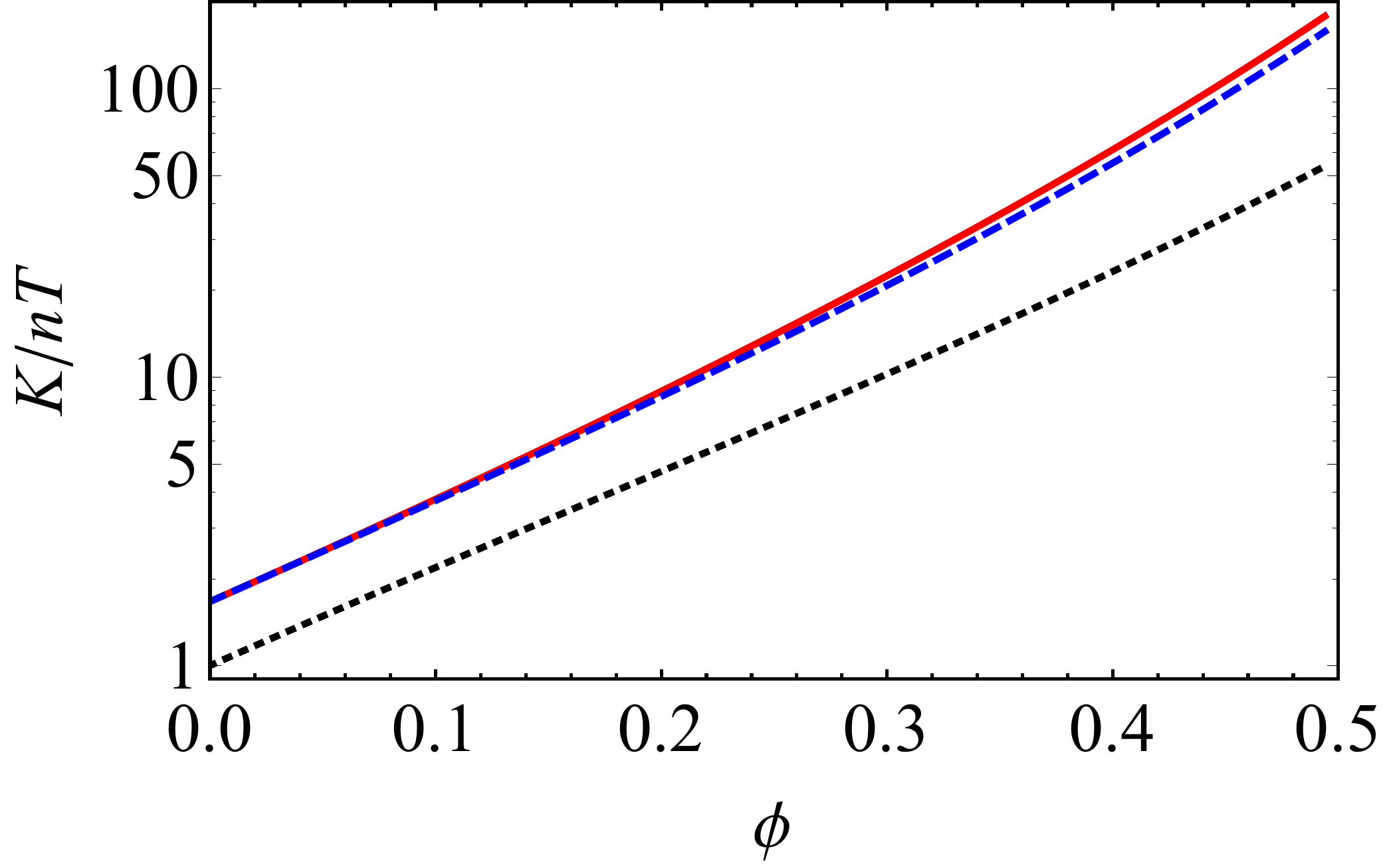}
\caption{Reduced bulk modulus of a HS fluid versus the packing fraction $\phi$. The red solid curve is calculated using Eq.~(\ref{K}). The blue dashed curve corresponds to the adiabatic bulk modulus of Eq.~(\ref{Adiabatic}). The black dotted curve shows the isothermal bulk modulus for comparison.}
\label{Fig1}
\end{figure}

Figure~\ref{Fig1} compares the elastic moduli defined by Eqs.~(\ref{K}) and (\ref{Adiabatic}). We see that
$K$ and $K_S$ are quite close. The relative deviation increases with $\phi$ and reaches $\simeq 10\%$ near the fluid-solid phase transition. Both quantities are considerably larger than the isothermal bulk modulus $K_T=nT\left[Z(\phi)+\phi dZ(\phi)/d\phi\right]$ (black dotted curve in Fig.~\ref{Fig1}). The ratio $K_S/K_T$ increases monotonously from $5/3$ at $\phi\rightarrow 0$  (as it should be in the ideal gas) to $\simeq 3$ at $\phi\simeq 0.5$. Overall, the rather good quantitative agreement between $K$ and $K_S$ gives us confidence in the discussed approach. Some of its immediate consequences are discussed below.     

\section{Consequences}

The phase diagram of HS systems is relatively simple~\cite{RintoulPRL1996,TorquatoRMP2010}. The phase state is determined by a single parameter -- packing fraction $\phi$. The fluid branch starts at $\phi=0$ and continues up to the freezing point at $\phi_{f}\simeq 0.494$. Fluid and solid coexist in the range between the freezing and melting points, $\phi_m\simeq 0.545$. The maximum solid packing fraction is the close-packed fcc crystal with $\phi_{cp}\simeq 0.74$. There is also a metastable extension of the fluid phase (that can be regarded as glass) above the melting point, which can be visited under rapid compression. The most rapid compression likely leads to the random close-packing, or, according to the new concept the maximally random jammed (MRJ) state with $\phi\simeq 0.64$~\cite{TorquatoRMP2010,TorquatoPRL2000,KlumovPRB2011}.       

Miller's derivation assumes isotropic conditions, and hence his results are not applicable to the solid phase. They may be applicable to the amorphous glassy phase, but to analyze this quantitatively we would need an equation of state and derivative of the RDF at contact fundamentally different from those specified in Sec.~\ref{MillerRes}. Therefore, the calculations are presented for the packing fraction below that at the freezing point. On the other hand, recent investigation~\cite{BrykJCP2017} has demonstrated that the shear mode can only be supported in HS fluids at sufficiently high densities. Although the instantaneous shear modulus of Eq.~(\ref{G}) is finite at any density due to the presence of the kinetic term, it is apparently not a relevant quantity at low densities~\cite{BrazhkinJPCB2018}. Consequently, the minimum packing fraction is chosen $\phi\gtrsim 0.2$. For these reasons, numerical results will be presented for the range $0.2\leq \phi \leq 0.494$.

It is convenient to present the calculations in terms of the longitudinal and transverse elastic sound velocities. The longitudinal velocity $c_l$ is related to the instantaneous longitudinal modulus $M=K+(4/3)G$, while the transverse sound velocity $c_t$ is expressed using the shear modulus. These relations are~\cite{Stratt1997}  
\begin{equation} 
c_l^2=M/mn, \quad\quad c_t^2=G/mn,
\end{equation}
where $m$ is the HS mass. The adiabatic sound velocity is
\begin{equation}
c_s=K_S/mn,
\end{equation}
and from the results above this is close to the high-frequency (instantaneous) sound velocity, $c_{\infty}\simeq K/mn$. The characteristic velocity scale associated with the thermal motion of hard spheres is the thermal velocity $v_T=\sqrt{T/m}$. In the following all velocities are expressed in units of thermal velocity.      

\begin{figure}
\includegraphics[width=7.5cm]{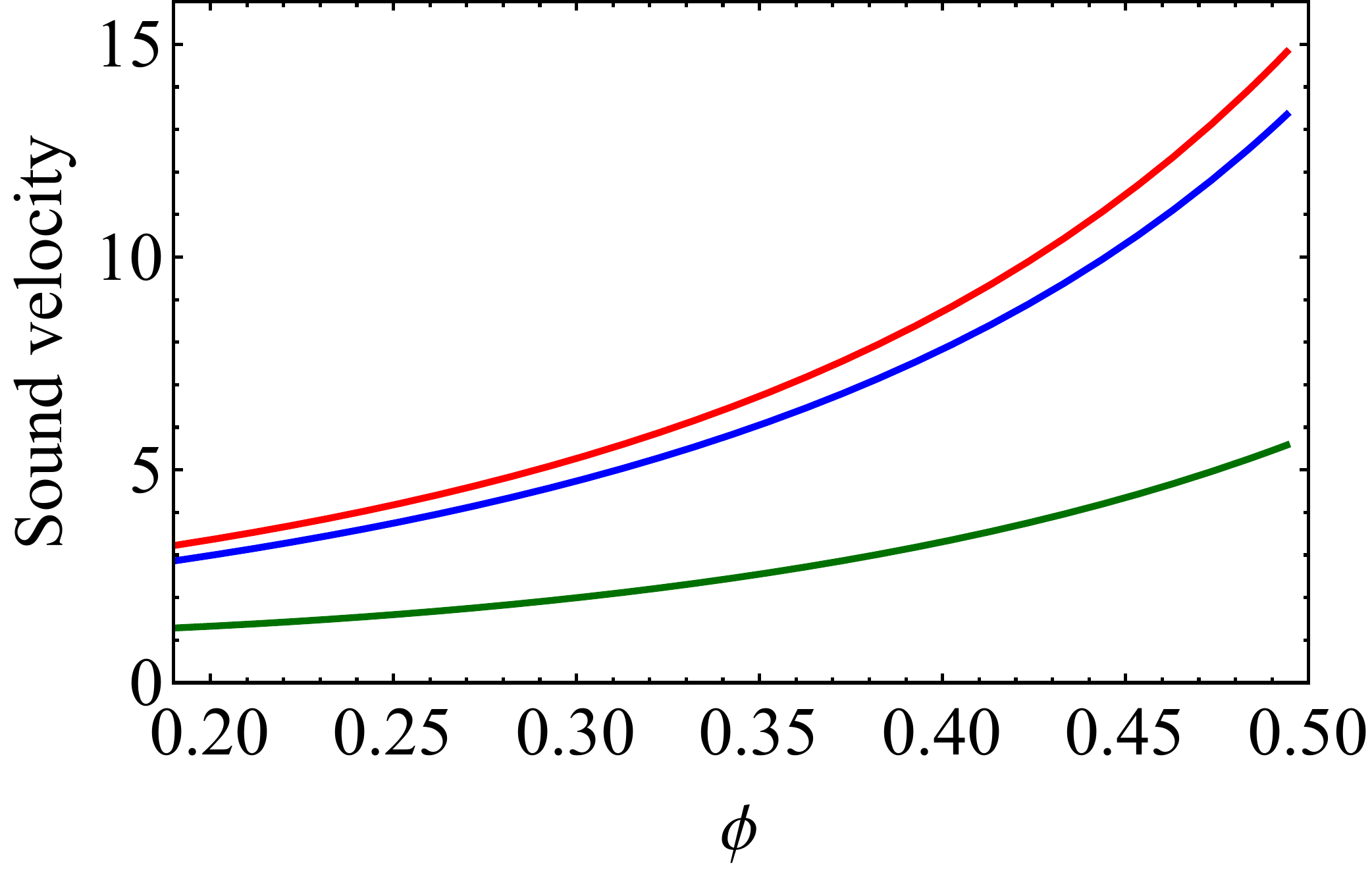}
\caption{Reduced sound velocities in a HS fluid versus the packing fraction $\phi$. Curves from top to bottom correspond to the longitudinal, instantaneous ($\simeq$adiabatic), and transverse velocity, respectively.}
\label{Fig2}
\end{figure}

Figure~\ref{Fig2} presents the calculation of acoustic velocities in HS fluids. The curves from top to bottom correspond to the longitudinal, instantaneous ($\simeq$ adiabatic), and transverse velocity, respectively. The reduced velocities increase monotonously on approaching the freezing point. At this point ($\phi= 0.494$) we get $c_l/v_T\simeq 14.8$, $c_s/v_T\simeq 12.6$, and $c_t/v_T\simeq 5.6$. The exact numbers can have some (relatively weak) dependence on the concrete form of the derivative of the RDF at contact and equation of state~\cite{KhrapakJCP2016}.      

\begin{figure}
\includegraphics[width=7.5cm]{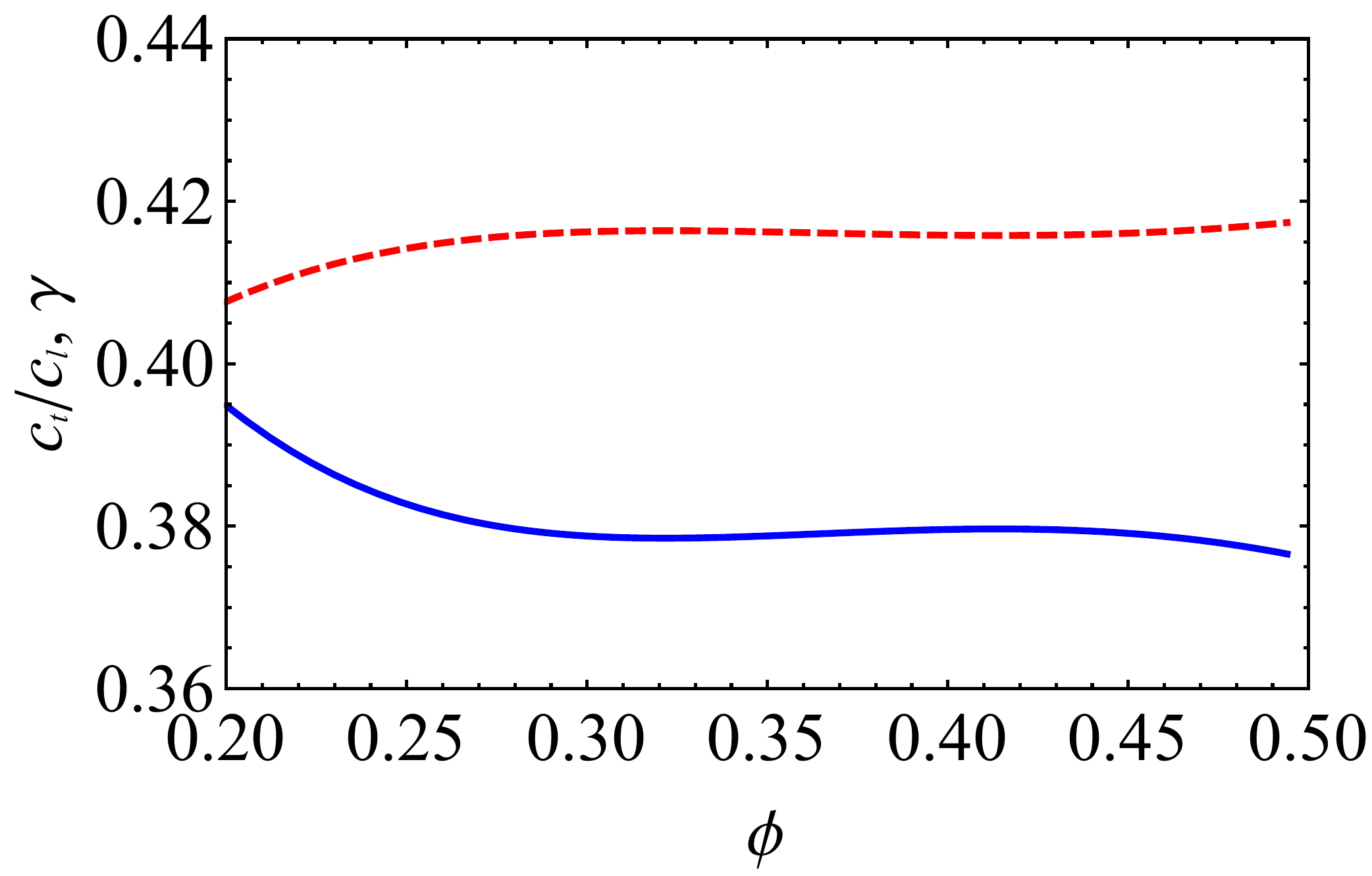}
\caption{Ratio of the transverse to the longitudinal sound velocity $c_t/c_l$ (blue solid curve) and Poisson's ratio (red dashed curve) of a HS fluid versus the packing fraction $\phi$. }
\label{Fig3}
\end{figure}

The ratio of the longitudinal and transverse sound velocities is plotted in Fig.~\ref{Fig3} (solid curve). It is relatively weakly dependent on $\phi$ with $c_t/c_l\simeq 0.38$ in the dense fluid regime. A quantity, directly related to the ratio of the longitudinal and transverse sound velocities, is Poisson's ratio~\cite{Greaves2011}, which can be expressed as
\begin{equation}
\gamma = \frac{1}{2}\frac{c_l^2-2c_t^2}{c_l^2-c_t^2}.
\end{equation} 
In the present context, $\gamma$ should be referred to as the infinite-frequency or instantaneous Poisson's ratio. 
This quantity is shown in Fig.~\ref{Fig3} by the dashed curve. It does not vary much in the dense fluid regime and numerically it is slightly above $0.4$ in the entire range of densities considered. Note that Poisson's ratio of about $0.4$ has been reported for an fcc HS crystal at thermodynamically unstable density corresponding to a fluid-solid coexistence~\cite{FrenkelPRL1987}.    

\begin{figure}
\includegraphics[width=7.5cm]{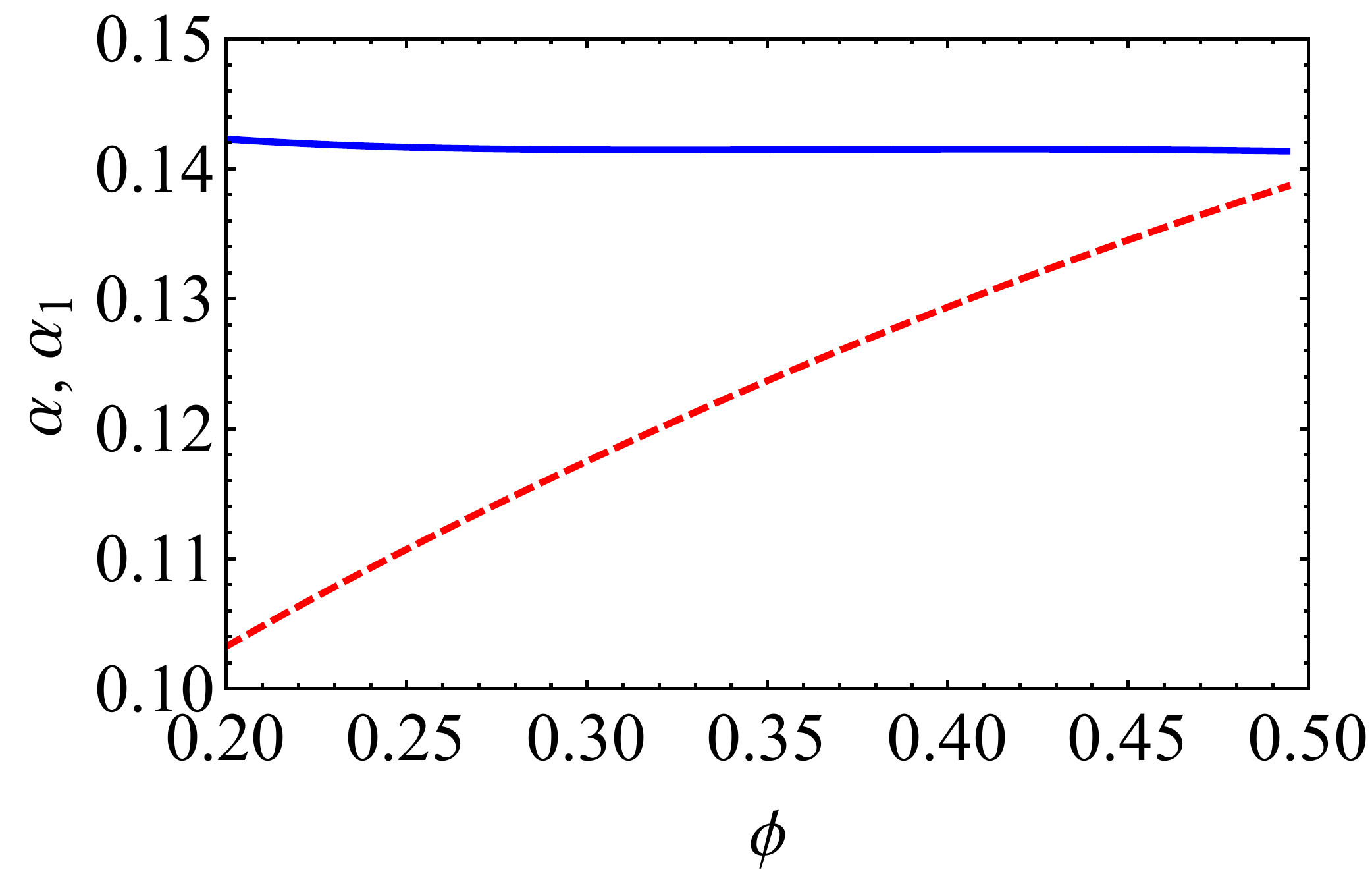}
\caption{Stokes-Einstein coefficient of a HS fluid versus the packing fraction $\phi$. The solid curve corresponds to the SE coefficient as defined by Eq.~(\ref{SE1}).  The dashed curve corresponds to an alternative formulation, Eq.~(\ref{SE2}). }
\label{Fig4}
\end{figure}

The ratio of sound velocities can also be important in the context of the Stokes-Einstein (SE) relation. For self-diffusion of atoms in simple pure fluids the SE relation takes the form~\cite{ZwanzigJCP1983,Balucani1990,OhtoriJCP2018,KhrapakAIPAdv2018,
CostigliolaJCP2019,KhrapakJCP2019}
\begin{equation}\label{SE1}
D\eta(\Delta/T)=\alpha,
\end{equation}
where $D$ is the self-diffusion coefficient, $\eta$ is the shear viscosity coefficient,  $\Delta=n^{-1/3}$ is the mean interparticle separation, and $\alpha$ is the SE coefficient. Moreover, the value of $\alpha$  can be related to the properties of
collective excitations~\cite{ZwanzigJCP1983}, and expressed in terms of the transverse-to-longitudinal sound velocity ratio, $\alpha\simeq 0.13(1+c_t^2/2c_l^2)$~\cite{KhrapakJCP2019}.  The expected dependence of the SE coefficient on $\phi$ for a dense HS fluid, resulting from this approximation, is plotted in Fig.~\ref{Fig4} (solid curve). This dependence is very weak: $\alpha$ is just above $0.14$ in the entire densities range considered. Recent MD simulations~\cite{OhtoriJCP2018} reported $\alpha$ in the range from $1/2\pi\simeq 0.159$ to $1/6\simeq 0.167$ for a similar range of $\phi$, which is somewhat above the theoretical expectations. It should be noted, however, that the effect of a finite particle number in simulations of the transport coefficients of HS fluids and the way how these coefficients are approaching the thermodynamic limit are not very thoroughly investigated (note, however, a very recent paper~\cite{HSNew}, where this thematics has been addressed). This can be one of the possible reasons behind the observed discrepancy. In an alternative form of the SE relation, the HS diameter $\sigma$ is used instead of the mean interparticle separation~\cite{Heyes2007}
\begin{equation}\label{SE2}
D\eta(\sigma/T)=\alpha_1.
\end{equation}  
The calculated coefficient $\alpha_1$ is shown in Fig.~\ref{Fig4}. It demonstrates a stronger dependence on the packing fraction and thus is less appropriate. The numerical value  $\alpha_1\simeq 0.14$ at freezing is again somewhat lower than that obtained in MD simulations ($\simeq 1/2\pi$)~\cite{Heyes2007,HeyesJPCM2007}.  

The generalized Cauchy identity relating the shear and bulk (or longitudinal) moduli of isotropic fluids is very well known~\cite{ZwanzigJCP1965,Schofield1966,FiorettoJCP2008}. It is obtained from expressions (\ref{G_}) and (\ref{K_}) by requiring cancellation of the terms with $\varphi''(r)$ and relating the rest to the pressure, which contains the term with $\varphi'(r)$ (see below). The results is

\begin{equation}\label{GenC}
M-3G = 2(P-nT).
\end{equation}  
At zero absolute temperature the right-hand-side of (\ref{GenC}) vanishes and we recover $M=3G$ or $K=\tfrac{5}{3}G$, known in the theory of elasticity of solids~\cite{LL_elasticity}. Originally, it was thought to be applicable to any isotropic fluid with two-body central interactions between the particles~\cite{ZwanzigJCP1965}. However, the derivation assumes continuous and differentiable potentials and hence does not apply directly to HS fluids. By comparing Eqs.~(\ref{G}) and (\ref{K}) we immediately see that Eq.~(\ref{GenC}) is not satisfied. Instead, Miller's result implies
\begin{equation}
M-3G = 2\left(P-nT\right)+\frac{2}{3}\left(\frac{P^2}{nT}-nT\right).
\end{equation}      
The second term on the right dominates at finite temperatures and high densities. At zero temperature, $M=3G$ is again recovered.   

\section{Discussion}

Let us return to the important problem of the divergence of conventional expressions for the high-frequency elastic moduli on approaching the HS interaction limit. By following a simple conventional derivation of the expression for the excess bulk modulus (which incorrectly leads to the diverging modulus) it is demonstrated where this (unphysical) divergence emerges from.   

For simplicity, we start with the one-dimensional (1D) situation. The starting point is the virial expression for the pressure~\cite{Tonks1936}
\begin{equation}
P=nT - (1/V)\sum r\varphi'(r),
\end{equation}
where the summation extends over all pairs of particles. We use the conventional 3D notation, one should bear in mind than in 1D case the role of pressure is played by the force length product, volume is the length,  density is inversely proportional to the length, etc. The virial expression can be expressed in the integral form using the RDF $g(r)$ and summing over particles~\cite{HansenBook}. The excess pressure (associated with the interactions between the particles is)
\begin{equation}\label{Pex1}
P_{\rm ex}= -n^2\int_0^{\infty}dr r\varphi'(r)g(r).
\end{equation}
Here the factor $\tfrac{1}{2}$ which should be present to avoid double summation is canceled when we substitute $\int_{-\infty}^{\infty}(...)dr$ by  $2\int_{0}^{\infty}(...)dr$. The excess bulk modulus  $K_{\rm ex}=n(\partial P_{\rm ex}/\partial n)$ is
\begin{equation}\label{K1}
K_{\rm ex} = 2P_{\rm ex}-n^3\int_0^{\infty}dr r \varphi'(r)\frac{\partial g(r)}{\partial n},
\end{equation} 
where the second term on the right-hand side includes the implicit density dependence of the RDF.  In general it depends on the particular thermodynamic process considered.  In the high-frequency (instantaneous) limit no relaxation is allowed and the density change occurs without any rearrangement of the particles~\cite{SingwiPRA1970}. This implies that the RDF, scaled by the interparticle separation, is a universal function, $g(rn)={\rm const}$. From the thermodynamic perspective the process considered occurs at a constant excess entropy (and this may explain why the instantaneous bulk modulus is usually close to the adiabatic one). In the context of isomorph concept, the  considered path is along an isomorph where structure and dynamics in properly reduced units
are invariant to a good approximation~\cite{DyreJPCB2014}. The requirement for the RDF is $g(r+\delta r; n+\delta n)=g(r;n)$ under $rn={\rm const}$, which immediately results in   
\begin{equation}\label{RDF1D}
\frac{\partial g (r; n)}{\partial n}=\frac{r}{n}\frac{\partial g(r; n)}{\partial r}.
\end{equation} 
Substituting this into Eq.~(\ref{K1}), integrating by parts, and using Eq.~(\ref{Pex1}) we finally obtain
\begin{equation}\label{K3D}
K_{\rm ex}=n^2\int_0^{\infty}dr r^2g(r)\varphi''(r).
\end{equation} 
Consider now the IPL interaction potential $\varphi(r)=\epsilon(\sigma/r)^{\ell}$, where $\epsilon$ is the energy scale. It approximates the HS potential as $\ell^{-1}\rightarrow 0$. Simple algebra allows us to relate $K_{\rm ex}$ and $P_{\rm ex}$ in this special case:
\begin{equation}
K_{\rm ex}=(\ell+1)P_{\rm ex}.
\end{equation}
The pressure in the HS limit is known {\it exactly} in 1D case. The Tonks results is~\cite{Tonks1936}
\begin{equation}
P_{\rm ex}=nT\frac{\phi}{1-\phi}.
\end{equation}
It only diverges at the highest packing fraction $\phi=1$, otherwise remaining finite. From this we conclude that the 1D bulk modulus exhibits  a divergence $\propto \ell$ as $\ell^{-1}\rightarrow 0$. The origin of this paradoxical behaviour can be immediately identified. It is the assumption of no structural rearrangement [independence of $g(rn)$ of $n$] that causes problems. While this is a very well justified assumption for sufficiently soft interactions, it is clearly not applicable to HS-like interactions, because an intrinsic length scale -- the hard sphere diameter  (or the hard rod length in 1D) -- emerges. Thus, the divergence of the high frequency elastic moduli in the limit of very steep interactions appears artificial. Rather, the conventional expressions must not be applied in this limit.   

The derivation is easily generalized to the 3D situation. We begin with the 3D virial expression~\cite{Tonks1936}
\begin{equation}
P=nT - (1/3V)\sum r\varphi'(r),
\end{equation}
to get
\begin{equation}\label{Pex2}
P_{\rm ex}= -\frac{2\pi n^2}{3}\int_0^{\infty}r^3\varphi'(r)g(r) d r.
\end{equation}
Differentiating (\ref{Pex2}) with respect to $n$ gives
\begin{equation}\label{dK}
K_{\rm ex}= 2P_{\rm ex}-\frac{2 \pi n^3}{3}\int_0^{\infty}r^3\varphi'(r)\frac{\partial g(r)}{\partial n} d r. 
\end{equation} 
The requirement of no rearrangement and the constancy of $g(rn^{1/3})$ in 3D results in~\cite{SingwiPRA1970,KhrapakPoP2016_Onset} 
\begin{equation}\label{RDF3D}
\frac{\partial g (r; n)}{\partial n}=\frac{r}{3n}\frac{\partial g(r; n)}{\partial r}.
\end{equation} 
Substituting this into Eq.~(\ref{dK}) and integrating by parts we immediately obtain
\begin{equation}\label{K3D}
K_{\rm ex}=\frac{2\pi n^2}{9}\int_0^{\infty}dr r^3g(r)\left[r\varphi''(r)-2\varphi'(r)\right].
\end{equation} 
This coincides with the excess part of Eq.~(\ref{K_}). For the IPL interaction we get
\begin{equation}
K_{\rm ex}=\frac{\ell+3}{3}P_{\rm ex}.
\end{equation}
The excess pressure is finite (and positive) for all $\ell>3$ and smoothly approaches the HS limiting value as $\ell^{-1}\rightarrow 0$~\cite{AgrawalMolPhys1995}. This implies that the bulk modulus exhibits again a divergence $\propto \ell$ as $\ell^{-1}\rightarrow 0$. The same is true for the shear modulus. This was recognized quite early~\cite{Frisch1966} and has remained a controversial issue since then. For different opinions see for instance Refs.~\cite{MillerJCP1969,HeyesJCP1994,HeyesMolPhys1998,PowlesMolPhys2000,RickayzenJCP2003,DuftyGraMat2011}. 
On the other hand, since the structural and thermodynamic properties are approaching smoothly and continuously the HS limit~\cite{AgrawalMolPhys1995,KhrapakJCP2011,JoverJCP2012,YurchenkoJPCM2016}, similar behavior should be expected from the elastic properties~\cite{DuftyMolPhys2004}.  

From the derivation above we see where the problem is. Exactly as in 1D case, the assumption of no structural rearrangement [independence of $g(rn^{1/3})$ of $n$] is not consistent with the intrinsic length scale in the system -- the hard sphere diameter. The conventional expressions for elastic moduli are just meaningless in the HS limit. 

We can formulate our result a bit differently. All the reduced properties of HS systems are uniquely determined by a single parameter -- the packing fraction. Therefore, the thermodynamic path that conserves the RDF (excess entropy) upon changing the packing fraction is simply not possible at equilibrium. This makes convenient formulas inapplicable.  

In our previous study of collective motion in IPL melts~\cite{KhrapakSciRep2017}, it was observed that the conventional expressions for the elastic moduli are apparently applicable down to $\ell^{-1}\sim 0.04$. The question of how the HS limit is approached for $0\lesssim\ell^{-1}\lesssim 0.04$ remains unsolved and we hope to address it in future work. 

\section{Conclusion}

Main conclusions can be formulated as follows. The elastic moduli of HS fluids that are responsible for sound velocities and related elastic properties are finite. Expressions derived by Miller~\cite{MillerJCP1969} have been numerically analyzed and some consequences have been pointed out. The origin behind the unphysical divergence of the conventional expressions for the instantaneous elastic moduli when approaching the HS limit has been identified and discussed. Using the IPL repulsive potential ($\propto r^{-\ell}$) as a reference example, we suggest that the conventional expressions can only be applied in the sufficiently soft interactions regime,  $\ell\lesssim 25$. Fortunately, this range includes most of the interactions existing in real world. Nevertheless, it would be interesting to investigate how the HS limit is reached at $\ell\gtrsim 25$. The results presented in this article can serve as a first step towards this goal.              

The question about what can be the frequency response of a HS fluid when the frequency of external perturbation increases above all frequencies relevant to the system has not been considered. There are theories that predict frequency divergence of the shear modulus as $\propto\sqrt{\omega}$ in this truly infinite-frequency limit~\cite{CichockiPRA1991}. This point is beyond the scope of the present work.

\acknowledgments
I would like to thank Bruce Miller for the correspondence and useful suggestions and Mierk Schwabe for careful reading of the manuscript.

\appendix*

\section{Derivative of the RDF at contact}

According to the derivation by Thiele~\cite{ThieleJCP1963} the RDF $g(x)$ can be related to the function $h(x)$, such that
$g(x)=h(x)/x$ for $x>1$. At contact the first three derivatives of $h(x)$ are continuous and $g'(1)=h'(1)-h(1)$. The function $h(x)$ in the range $0<x<1$ is given by a simple polynomial form
\begin{displaymath}
h(x)=ax+bx^2+cx^4,
\end{displaymath}   
where
\begin{displaymath}
\begin{split}
a=& \frac{(2\phi+1)^2}{(1-\phi)^4},\\
b=& \frac{-(12\phi+12\phi^2+3\phi^3)}{2(1-\phi)^4},\\
c=& \frac{\phi(2\phi+1)^2}{2(1-\phi)^4}.
\end{split}
\end{displaymath}
This results in $g'(1)=b+3c$ and, after some simple algebra, Eq.~(\ref{RDF}) is obtained.

\bibliographystyle{aipnum4-1}
\bibliography{HS_References}

\end{document}